# VCube-PS: A Causal Broadcast Topic-based Publish/Subscribe System

João Paulo de Araujo[a,*], Luciana Arantes[a], Elias P. Duarte Jr.[b], Luiz A. Rodrigues[c], Pierre Sens[a]

[a]*Sorbonne Université, CNRS, Inria, LIP6 – 4, place Jussieu, 75252, Paris Cedex 5, France*

[b]*Federal University of Paraná (UFPR) – UFPR/Depto. Informática, Caixa Postal 19018, Curitiba, PR, 81531-980, Brazil*

[c]*Western Paraná State University (Unioeste) – R. Universitária, 2069, Universitário, Cascavel, PR, 85819-110, Brazil*

**Abstract**

In this work we present *VCube-PS*, a topic-based Publish/Subscribe system built on the top of a virtual hypercube-like topology. Membership information and published messages are broadcast to subscribers (members) of a topic group over dynamically built spanning trees rooted at the publisher. For a given topic, the delivery of published messages respects the causal order. *VCube-PS* was implemented on the PeerSim simulator, and experiments are reported including a comparison with the traditional Publish/Subscribe approach that employs a single rooted static spanning-tree for message distribution. Results confirm the efficiency of *VCube-PS* in terms of scalability, latency, number and size of messages.

*Keywords:*
Publish/Subscribe, Topic-based Pub/Sub, Causal Broadcast, Distributed Spanning Trees, Hypercube-like Topologies

---

[*]Corresponding author

  *Email addresses:* `joao.araujo@lip6.fr` (João Paulo de Araujo), `luciana.arantes@lip6.fr` (Luciana Arantes), `elias@inf.ufpr.br` (Elias P. Duarte Jr.), `luiz.rodrigues@unioeste.br` (Luiz A. Rodrigues), `pierre.sens@lip6.fr` (Pierre Sens)

1. **Introduction**

Publish/Subscribe (Pub/Sub) systems consist of a set of *publishers* which are distributed nodes that publish messages that are consumed by *subscribers*. The communication between publishers and subscribers is conducted on an overlay infrastructure, which is generally composed by a set of nodes that organize themselves for ensuring the delivery of published messages to all (and preferably only) subscribers interested in those messages. Hence, publishers and subscribers exchange information asynchronously, without interacting directly [1, 2]. They might even not know each other.

In topic-based Pub/Sub system, a subscriber can register its interests in one or more topics, and then it receives all published messages related to these topics (e.g., Scribe [3], Bayeux [4], DYNATOPS [5], Dynamoth [6], Magnet [7] and DRScribe [8]). The advantages of topic-based Pub/Sub systems when compared to content-based (see Section 5) are mainly that messages can be statically grouped into topics, the diffusion of messages to subscribers is usually based on multicast groups, and the interface offered to the user is simple. The topic approach is widely used by popular applications including Twitter and Firebase/Google Cloud Messaging, IBM MQ, distributed multiplayer online games, chat systems, and mobile device notification frameworks.

Many topic-based Pub/Sub systems found in the literature are based on per topic broadcast trees built over P2P DHTs [3, 4, 6, 8]. A single multicast tree is associated to each topic composed by both subscribers (resp., brokers) and forwarders, i.e., non-subscribers (resp., non-brokers) of the topic. Therefore, all publish messages related to a topic are broadcast through the same tree. In this work, we call these systems *SRPT* (*Single Root Per Topic*). As they are built over P2P DHTs, they are scalable in terms of the number of subscribers. On the other hand, maintenance of the one single tree per topic can be costly, particularly when the membership of the system changes [3, 5]. *SRPT* employs multiple forwarders, which are nodes that do not deliver the messages themselves but are employed in the dissemination. Forwards induce higher latency and, in case of a high number of simultaneous publications of a single topic, the root of the tree presents contention problems, becoming a performance bottleneck.

In [9], the authors show that in applications like Twitter, most of the publications are concentrated in few topics: roughly 83% of the analyzed topics have up to 5 published messages and only 0.15% of the topics ("hot topics") are related to more than 1,000 publishing messages. An example of such applications is multiplayer online combat games where locations are mapped to topics [10, 11]. During the game, players move towards the same location increasing the publishing load for the topic corresponding to the location, i.e., the location becomes a "hot topic". We argue that *SRPT*-based Pub/Sub systems are not suitable to handle a high publishing load as they present root contention constraints.

We also claim that a topic-based Pub/Sub system must enforce, for a given topic, the causal order of published messages. For instance, in a discussion group, a question published on a group should never be delivered to any subscriber after an answer to that question which was also published in the same group, as the answer is causally related to the question. In other words, if a node publishes a message after it delivers another message, then no node delivers the latter after the former. To the best of our knowledge, only JEDI [12]) provides causal ordering of published messages.

Considering the above discussed points, we propose in this article *VCube-PS*, a non DHT Pub/Sub system, that ensures low latency, and load balancing for publishing messages. It also respects the causal delivery order of published messages to a same topic, which is a crucial feature for several types of Pub/Sub applications. In our system, a published message is sent to all subscribers of a topic by a broadcast protocol that creates a spanning tree composed just by the subscribers, whose root is the publisher. Hence, the root "hot topic" contention problem of *SRPT* does not exist in *VCube-PS* since there is no single root tree per topic, as each node that publishes a message becomes the root of the corresponding spanning tree. Broadcast trees are dynamically built on top of a virtual hypercube-like topology, called *VCube* [13], that presents logarithmic properties, thus providing scalability. Contrarily to *SRPT* Pub/Sub systems and thanks to *VCube*'s properties, both the construction and maintenance of spanning trees by *VCube-PS* have no overhead, even in the presence of subscriber membership changes. In other words, *VCube-PS* locally calculates to which nodes the messages need to be forwarded, without the need of routine tables. In the absence of churn, *VCube-PS* does not present forwarder nodes and in the presence of it, the latter are temporary.

We point out that, contrarily to *SRPT* systems, our target applications are mainly those that present "hot topics" (e.g. multiplayer combat games, company chat groups, etc.).

We implemented *VCube-PS* and two *SRPT*-like Pub/Sub systems on top of the PeerSim simulator [14]. One *SRPT* Pub/Sub is subscriber-based

(e.g., Scribe, Magnet, DRScribe) while the second one is broker-based (e.g. Dynatops). In Dynatops, subscribers are connected to brokers based on locality. Results confirm the advantages of using per-publisher dynamically built spanning trees in terms load balancing, latency, number and size of messages metrics.

The rest of the paper is organized as follows. Section 2 gives an overview of *VCube*. Section 3 presents *VCube-PS*'s algorithms to manage topics, order messages, as well as the specification of *VCube-PS*'s algorithms. Section 4 presents evaluation of results conducted on PeerSim simulator. Section 5 discusses related work and, finally, Section 6 concludes the paper.

## 2. VCube

In *VCube* [13], a node $i$ groups the other $N - 1$ nodes in $d = \log_2 N$ clusters forming a *d-VCube*, each cluster $s$ ($s = 1, .., d$) having $2^{s-1}$ nodes. The ordered list of nodes in each cluster $s$ is defined by function $c_{i,s}$ below, where $\oplus$ is the bitwise exclusive *or* operator (xor).

$$c_{i,s} = i \oplus 2^{s-1} \parallel c_{i \oplus 2^{s-1}, k} \mid k = 1, \ldots, s-1$$

This recursive function can be described as follows. Initially, the first, the neighbor of node $i$ in cluster $s$ is computed. The identifiers of these two nodes differ only in one bit, the bit that is set to one in $2^{s-1}$. Then, the remaining nodes in the cluster are nodes in clusters $1, \ldots, s-1$ of the hypercube neighbor, i.e., $c_{i \oplus 2^{s-1}, 1}, c_{i \oplus 2^{s-1}, 2}, \ldots, c_{i \oplus 2^{s-1}, s-1}$.

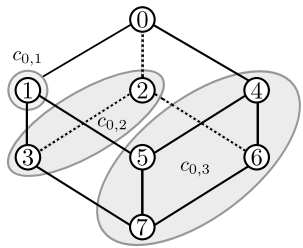

The $c_{i,s}$ table for 8 nodes

| s | $c_{0,s}$ | $c_{1,s}$ | $c_{2,s}$ | $c_{3,s}$ | $c_{4,s}$ | $c_{5,s}$ | $c_{6,s}$ | $c_{7,s}$ |
|---|---|---|---|---|---|---|---|---|
| 1 | 1 | 0 | 3 | 2 | 5 | 4 | 7 | 6 |
| 2 | 2 3 | 3 2 | 0 1 | 1 0 | 6 7 | 7 6 | 4 5 | 5 4 |
| 3 | 4 5 6 7 | 5 4 7 6 | 6 7 4 5 | 7 6 5 4 | 0 1 2 3 | 1 0 3 2 | 2 3 0 1 | 3 2 1 0 |

Figure 1: *VCube* hierarchical organization.

The table of Figure 1 contains, for $N = 8$, the composition of all $c_{i,s}$ of the 3-*VCube*. The same figure also shows node 0's hierarchical cluster-based logical organization in the 3-*VCube*.

## 3. VCube-PS: Publish/Subscribe System

In this section, we present the topic-based *VCube-PS* Pub/Sub system. We first describe the system model. Then, we describe causal order broadcast, the use of *causal barriers*, and the per-source FIFO reception order of *VCube-PS*. Finally, we give the algorithms that compose *VCube-PS*.

*3.1. System Model and Definitions*

We consider a distributed system composed of a finite set of $\Pi = \{0,..,N-1\}$ nodes with $N = 2^d$ nodes, $d > 0$. Each node has a unique identifier ($id$) and nodes communicate only by message passing. A user of the Pub/Sub system corresponds to a node. Nodes are organized in a logical hypercube.

Nodes communicate by sending and receiving messages. The network is fully connected: each pair of nodes is connected by a bidirectional point-to-point channel and there is no network partitioning. Nodes do not fail and links are reliable. Thus, messages exchanged between any two processes are never lost, corrupted nor duplicated. The system is asynchronous, i.e., relative processor speeds and message transmission delays are unbounded.

The *source* of a message is the node that broadcasts the message. We distinguish between the arrival of a message (*reception*) at a process and the event at which the message is given to the application/user (*delivery*). Only the latter respects the causal order of published messages.

*3.2. Causal and Per-source FIFO Reception Ordering*

For each topic, *VCube-PS* enforces the causal order of published messages, implementing, thus, causal broadcast. It also implicitly ensures that for a single publisher, nodes will receive messages in the order they were published.

*3.2.1. Causal Ordering*

For a given topic $t$, if a process publishes a message $m'$ after it has delivered a message $m$, then no process in the system will deliver $m$ after $m'$. Note that if a process $i$ never delivers $m'$ (i.e., $i$ leaves the topic before delivering $m'$) or delivers $m'$ but never delivers $m$ (i.e., $i$ was not subscribed to $t$ when $m$ was published), the causal order of published messages is not violated.

In order to implement the causal order of published messages, we apply *causal barriers* [15]. The key advantage of the *causal barrier* approach is that it does not enforce the causal order based on the identifiers of the nodes (per node vector) but by using direct message dependencies, which renders the algorithm more suitable for dealing with the node dynamics (subscriptions and unsubscriptions), in comparison to other vector clock-based implements of causal broadcast such as [16] or [17].

Let $m$ and $m'$ be two application messages published for topic $t$. Message $m$ immediately precedes $m'$ ($m \prec_{im} m'$) if (1) the publishing of $m$ causally precedes the publishing of $m'$ and (2) there exists no message $m''$ such that the publishing of $m$ causally precedes the publishing of $m''$, and the publishing of $m''$ causally precedes the publishing of $m'$. The *causal barrier* of $m$ ($cb_m$) consists of the set of messages that are immediate predecessors of $m$.

Figure 2 shows a distributed system with three nodes ($p_0$, $p_1$, and $p_2$) that have subscribed to the same topic $t$. Message $m_{s,t,c}$ is the message published by $s$ with sequence number $c$ for topic $t$. On the left, a timing diagram shows messages being published and delivered; the graph with message dependencies is shown on the right side. We can observe that the delivery of $m_{1,t,1}$ is conditioned by the delivery of $m_{0,t,1}$ ($m_{0,t,1} \prec_{im} m_{1,t,1}$) since $p_1$ delivered $m_{0,t,1}$ before publishing $m_{1,t,1}$, (i.e., $cb_{m_{1,t,1}} = \{m_{0,t,1}\}$). On the other hand, $m_{1,t,2}$ directly depends on $m_{2,t,1}$ and $m_{1,t,1}$ (i.e., $cb_{m_{1,t,2}} = \{m_{2,t,1}, m_{1,t,1}\}$). Note that since $m_{0,t,1}$ precedes $m_{1,t,1}$ that precedes $m_{1,t,2}$, $m_{0,t,1}$ is an indirect dependency of $m_{1,t,2}$, and was not included, therefore, in $cb_{m_{1,t,2}}$.

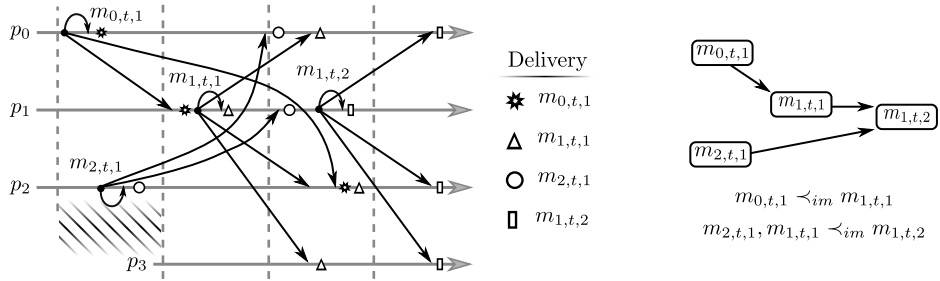

Figure 2: Example of causal barrier.

Now suppose that in the same system shown in Figure 2, $p_3$ subscribes to $t$ after messages $m_{2,t,1}$ was published to the other nodes, i.e., node $p_3$ did not take part in the spanning trees that broadcast $m_{2,t,1}$ and, consequently, in this case, node $p_3$ will neither receive nor deliver $m_{2,t,1}$. Hence, after having delivered $m_{1,t,1}$, $p_3$ can deliver $m_{1,t,2}$. Since nodes can dynamically subscribe to or unsubscribe from a topic in *VCube-PS*, our implementation of causal order must distinguish between the case in which a message will be delivered (e.g., $m_{1,t,1}$) from the one that it will never be delivered (e.g., $m_{2,t,1}$ by $p_3$). To this end, *VCube-PS* guarantees the following property on the FIFO order of messages published on a given topic.

*3.2.2. Per-source FIFO Reception Ordering*

Messages published by a same publisher are received by subscribers in the same order as they were produced. This order allows a subscriber of $t$ to know that it will never receive some messages previously published, i.e., if $m'_{s,t,c'}$ is the first message that node $i$ receives from $s$ on topic $t$ after it joined $t$'s group, $i$ will never receive $m_{s,t,c}$, $\forall c < c'$.

In *VCube-PS*, per-source FIFO reception order is ensured by the acknowledgment of published messages: a source node broadcasts a new message only after having received all the acknowledgments for the previous message it broadcast. Note that the per-source FIFO reception order is defined in regard to the reception of messages and not delivery, as in the traditional FIFO order definition.

*3.3. Algorithms*

This section presents *VCube-PS*'s algorithms. *VCube-PS* is based on *VCube*, which organizes nodes in a logical hypercube-like topology. Note that in *VCube-PS* nodes do not fail, *VCube-PS* exploits *VCube*'s organization but not its failure detection functionality.

**Types of Messages, Local Variables and Auxiliary functions**: Each message $m$ is uniquely identified by the source ($s$) and a sequence counter ($c$). It also carries information about the topic $t$. Messages can be of type SUB (subscription), UNS (unsubscribe), PUB (publication), and ACK (acknowledge). The value of the *data* field depends on the *type* of the message: for SUB and UNS messages, it holds no information while for PUB or ACK messages, it respectively holds the published message itself plus causal dependencies (*causal barrier*). $MAX\_TOPICS$ is a constant value that limits how many topics the system can support.

The following local variables are kept by every node $i$:

- *counter*: it is a local counter of node $i$ which is incremented at every subscription, unsubscription, or publishing of a message by node $i$;

- $br\_queue[MAX\_TOPICS]$: each $br\_queue[t]$ is a set of pending messages (PUB, SUB, or UNS) related to the topic $t$ waiting to be broadcast;

- $view[MAX\_TOPICS]$: set of the latest subscription and unsubscription operations of which node $i$ is aware. Each entry $view[t]$ has format $\langle n, o, rc \rangle$ where $n$ is the identity of the node that has joined or left the topic $t$; $o$ is equal to SUB or UNS and $rc$ stores the value of the counter of $n$ at the moment the subscription or unsubscription took place;

- $causal\_barrier[MAX\_TOPICS]$: each $causal\_barrier[t]$ keeps information on all messages that are predecessors of the next message that will be published by node $i$ for topic $t$; the causal barrier consists thus of a set of message identifiers of format $\langle s, c \rangle$ (source and sequence counter).

- $acks$: set of pending ACK messages for which $i$ waits confirmation. For each message propagation to its $nb$ children in the spanning tree of a message $m$ identified by $\langle s, t, c \rangle$ received from $j$, $i$ adds the element $\langle j, nb, \langle s, t, c, mem \rangle \rangle$ to the $acks$ set. The set $mem$ gathers membership information sent by ACK messages;

- $msgs$: set of messages that are being temporarily kept by $i$ because they have not been delivered yet. Upon delivering $m$, identified by $\langle s, t, c \rangle$, the latter can be removed from $msgs$;

- $not\_delvs[MAX\_TOPICS]$: each $not\_delvs[t]$ contains a set of messages received by node $i$ for topic $t$ and not yet delivered because their respective *causal barrier* has not been satisfied. Each element has format $\langle s, c, cb \rangle$ where $s$ is the identity of the source node that broadcast the message whose counter is $c$, and $cb$ corresponds to the *causal barrier* of the message.

- $last\_delvs[MAX\_TOPICS]$: each $last\_delvs[t]$ keeps the identifiers of the last message from each publisher node delivered by node $i$ for topic $t$. Each element of the set is the tuple $\langle s, c \rangle$ where $s$ is the source identity of the message whose counter is $c$;

- $first\_rec[MAX\_TOPICS]$: each $first\_rec[t]$ keeps the identifiers of the first message received from each publisher for a topic $t$. Each element of the set is a tuple $\langle s, c \rangle$.

In the algorithms, the symbol $\perp$ represents a *null* element while the underscore (_) is used to indicate *any* element.

We have defined two auxiliary functions that exploit *VCube* organization and are used to dynamically build broadcast spanning trees:

- CLUSTER$(i, j)$: the function returns the index $s$ of the cluster of node $i$ that contains node $j$, $(1 \leq s \leq \log_2 N)$. For instance, in Figure 1, CLUSTER$(0, 1) = 1$, CLUSTER$(0, 2) =$ CLUSTER$(0, 3) = 2$, and CLUSTER$(0, 4) =$ CLUSTER$(0, 5) =$ CLUSTER$(0, 6) =$ CLUSTER$(0, 7) = 3$.
- CHILDREN$(i, t, h)$: returns a set with all nodes virtually connected to node $i$. A child of $i$ is the first node of cluster $c_{i,s}$ which is also a subscriber of topic $t$; or the first node in $c_{i,s}$ in case of no topic ($t = $ '$*$'). The parameter $h$ can range from 1 to $\log_2 N$. If $h = \log_2 N$, the result set contains the $i$'s children where each child is in $c_{i,s}, s = 1, .., \log_2 N$. For any other value of $h < \log_2 N$, the function returns only a subset of $i$'s children, i.e., those children whose respective cluster number $s$ is smaller or equal to $h$ ($s \leq h$) For instance, in Figure 1, if $t = $ '$*$', CHILDREN$(0, *, 3) = \{1, 2, 4\}$, CHILDREN$(0, *, 2) = \{1, 2\}$, and CHILDREN$(4, *, 2) = \{5, 6\}$. On the other hand, if only nodes 0, 3, and 4 have joined topic $t_1$, CHILDREN$(0, t_1, 3) = \{3, 4\}$ and CHILDREN$(4, t_1, 2) = \emptyset$.

**Application (User Interface) functions**: *VCube-PS* offers an interface consisting of functions SUBSCRIBE$(t)$, UNSUBSCRIBE$(t)$, and PUBLISH$(t, m)$, all presented in Algorithm 1. A node can publish a message related to a topic if it is currently a subscriber of this topic. These functions generate messages of types SUB, UNS, or PUB, respectively, which are sent to all nodes, in case of subscription, or all subscribers of topic $t$, otherwise.

**Propagation of a Message**: When node $i$ invokes one of the application functions (Algorithm 1) for topic $t$, the procedure CO_BROADCAST (line 5 of Algorithm 2) is called, generating a new message of the corresponding type (PUB, SUB, or UNS) which is inserted in the queue of $t$. Then, a task related to $t$ (Task $START\_MSG\_PROPAGATION$) continuously removes the first message from this queue and starts the broadcast. The next message is removed from the queue only after the reception of acknowledge (message ACK) from all current subscribers (per-source FIFO reception order) to whom node $i$ sent the previous message (line 31). The task associated with $t$ is created when node $i$ becomes a new subscriber of the group of topic $t$ (line 11).

Task $START\_MSG\_PROPAGATION$ for topic $t$ starts the propagation of $m$, the first message removed from the queue (line 15), by dynamically building a hierarchical spanning tree, rooted at $i$, composed by the nodes which are either the subscribers of $t$, in case of messages of type UNS or PUB or by all nodes, in case of messages of type SUB (lines 23-28). For this

**Algorithm 1** Functions offered as the interface to the application: node $i$

1: **Init**
2:     $counter \leftarrow 0$
3:     $\forall t \in MAX\_TOPICS : view[t] \leftarrow \emptyset$

4: **function** SUBSCRIBE(topic $t$)
5:     **if** $\langle i, SUB, \_ \rangle \notin view[t]$ **then**
6:         $view[t] \leftarrow \{\langle i, SUB, counter \rangle\}$
7:         CO_BROADCAST($SUB, t, \_$)
8:         **return** OK
9:     **return** NOK

10: **function** UNSUBSCRIBE(topic $t$)
11:     **if** $\langle i, SUB, \_ \rangle \in view[t]$ **then**
12:         $view[t] \leftarrow view[t] \setminus \{\langle i, SUB, \_ \rangle\}$     ▷ removes subscription for $t$
13:         CO_BROADCAST($UNS, t, \_$)
14:         **return** OK
15:     **return** NOK

16: **function** PUBLISH(topic $t$, message $data$)
17:     **if** $\langle i, SUB, \_ \rangle \in view[t]$ **then**     ▷ only subscribers of t can publish at t
18:         CO_BROADCAST($PUB, t, data$)
19:         **return** OK
20:     **return** NOK

purpose, $i$ calls function CHILDREN($i, t, \log_2 N$) which renders, for PUB and UNS messages, the set of the first subscriber nodes of $t$ for each of its clusters (line 26) or the first node of each of $i$'s clusters (line 24) in the case of a $SUB$ message ($t = $ ' $*$ '). These nodes become $i$'s children in the spanning tree and $m$ is sent to them. Upon the reception of $m$ from a node $j$, by calling function CLUSTER($i, j$) (line 42 or 44 depending on the type of message), every child of node $i$'s sends $m$ to its own children in the $s - 1$ clusters, in relation to topic $t$ and the cluster $s$ of $i$ to which $j$ belongs, i.e., $c_{i,s}$. These nodes then become $j$'s children, and so on.

For instance, consider the left side of Figure 3, that all nodes are subscribers of $t_1$, and that node $p_0$, subscriber of $t_1$, wants to publish a message $m_0$ related to $t_1$ (PUB messages). $p_0$ is the root of the respective spanning tree: $m_0$ will be sent to the $\log_2 N = 3$ children of $p_0$ (CHILDREN($0, t_1, 3$) = $\{1, 2, 4\}$). Upon the reception of message $m_0$, $p_1$ does not forward it since CHILDREN($1, t_1, 0$) = $\emptyset$, while $p_2$ forwards it to its child $p_3$, the first subscriber of cluster $c_{2,1}$ (CHILDREN($2, t_1, 1$) = $\{3\}$). When $p_3$ receives $m_0$, as CHILDREN($3, t_1, 0$) = $\emptyset$, $p_3$ does not forward $m_0$ to any node. However, in

**Algorithm 2** Causal broadcast algorithm and delivery executed by node $i$

---
1: **Init**
2: $\forall t \in MAX\_TOPICS$: $view[t] \leftarrow \emptyset$; $first\_rec[t] \leftarrow \emptyset$; $not\_delvs[t] \leftarrow \emptyset$; $delv[t] \leftarrow \emptyset$; $br\_queue[t] \leftarrow \emptyset$
3: $msg \leftarrow \emptyset$
4: **create task** $HANDLE\_RECEIVED\_MSG$

5: **procedure** CO_BROADCAST(message_type $type$, topic $t$, message $data$)
6:     NEW($m$)
7:     $m.type \leftarrow type$; $m.s \leftarrow i$; $m.t \leftarrow t$
8:     $m.c \leftarrow counter$; $m.data \leftarrow data$
9:     $counter \leftarrow counter + 1$
10:     **if** $type = SUB$ **then**
11:         **create task** $START\_MSG\_PROPAGATION(t)$
12:     $br\_queue[t].insert(m)$

13: **Task** $START\_MSG\_PROPAGATION$(topic $t$)
14: **loop**
15:     $m \leftarrow br\_queue[t].first()$                     ▷ block if queue is empty
16:     **if** $m.type = PUB$ **then**
17:         **if** $\langle i, \_\rangle \notin first\_rec[t]$ **then**
18:             $first\_rec[t] \leftarrow first\_rec[t] \cup \{\langle i, m.c\rangle\}$
19:         CO_DELIVER($m$)
20:         $last\_delvs[t] \leftarrow last\_delvs[t] \setminus \{\langle i, \_\rangle\} \cup \{\langle i, m.c\rangle\}$
21:         $m.cb \leftarrow causal\_barrier[t]$
22:         $causal\_barrier[t] \leftarrow \{\langle i, m.c\rangle\}$
23:     **if** $m.type = SUB$ **then**
24:         $chd \leftarrow$ CHILDREN($i, *, \log_2 N$)
25:     **else**
26:         $chd \leftarrow$ CHILDREN($i, t, \log_2 N$)
27:     **for all** $k \in chd$ **do**
28:         SEND($m$) **to** $p_k$
29:     **if** $chd \neq \emptyset$ **then**
30:         $acks \leftarrow acks \cup \{\langle \bot, \#(chd), \langle i, t, m.c, \emptyset\rangle\rangle\}$
31:     **wait until** ($acks \cap \{\langle \bot, \_, \langle m.s, m.t, m.c, \_\rangle\rangle\} = \emptyset$)
32:     **if** $m.type = UNS$ **then**
33:         $msg \leftarrow msg \setminus \{m \mid m.t = t\}$ $not\_delvs[t] \leftarrow \emptyset$
34:         $first\_rec[t] \leftarrow \emptyset$; $delv[t] \leftarrow \emptyset$
35:         **if** $br\_queue[t] = \emptyset$ **then**
36:             exit

```
37: Task HANDLE_RECEIVED_MSG
38: loop
39:     upon receive m from p_j                                          ▷ block if no message
40:         if m.type ≠ ACK then
41:             if m.type = SUB then
42:                 chd ← CHILDREN(i, *, CLUSTER(i,j) − 1)
43:             else
44:                 chd ← CHILDREN(i, m.t, CLUSTER(i,j) − 1)
45:             if chd = ∅ then                                          ▷ leaf node
46:                 NEW(m′)
47:                 m′.type ← ACK;  m′.s ← m.s;  m′.t ← m.t
48:                 m′.c ← m.c;  m.data ← ∅
49:                 SENDACKS(j, m′)
50:             else                                                     ▷ propagate m
51:                 acks ← acks ∪ {⟨j, #(chd), ⟨m.s, m.t, m.c, ∅⟩⟩}
52:                 for all k ∈ chd do
53:                     SEND(m) to p_k
54:         else                                                         ▷ m.type = ACK
55:             k, nb, mem ← k′, nb′, mem′ : ⟨k′, nb′, ⟨m.s, m.c, m.t, mem′⟩⟩ ∈ acks
56:             acks ← acks ∖ ⟨k, nb, ⟨m.s, m.c, m.t, mem⟩⟩
57:             m.data ← m.data ∪ mem
58:             if nb > 1 then
59:                 acks ← acks ∪ ⟨k, nb − 1, ⟨m.s, m.c, m.t, m.data⟩⟩
60:             else if k ≠ ⊥ then                                       ▷ All pending ACKs were received
61:                 SENDACKS(k, m)

62:         if ⟨i, SUB, _⟩ ∈ view[m.t] then                              ▷ i is subscribed to m.t
63:             if m.type = PUB then
64:                 if (∄⟨m.s, _⟩ ∈ first_rec[m.t]) then
65:                     first_rec[m.t] ← first_rec[m.t] ∪ {⟨m.s, m.c⟩}
66:                 not_delvs[m.t] ← not_delvs[m.t] ∪ {⟨m.s, m.c, m.cb⟩}
67:                 msgs ← msgs ∪ {m}
68:                 CHECKDELIVERY(m.t)                                   ▷ received messages may be delivered
69:             else if m.type = ACK then
70:                 view[m.t] ← UPDATE(view[m.t], m.data)
71:             else                                                     ▷ SUB or UNS message
72:                 view[m.t] ← UPDATE(view[m.t], {⟨m.s, m.type, m.c⟩})
73:                 if m.type = UNS then
74:                     first_rec[m.t] ← first_rec[m.t] ∖ {⟨m.s, _⟩}

75: function UPDATE(set_1, set_2)
76:     for all ⟨n_1, _, rc_1⟩ ∈ set_1 do
77:         if (∃ ⟨n_1, _, rc_2⟩ ∈ set_2) then
78:             if rc_2 > rc_1 then
79:                 set_1 ← set_1 ∖ {⟨n_1, _, rc_1⟩}
80:             else
81:                 set_2 ← set_2 ∖ {⟨n_1, _, rc_2⟩}
82:     return set_1 ∪ set_2
```

```
83: procedure CHECKDELIVERY(topic t)
84:     while (∃ ⟨s, c, cb⟩ ∈ not_delvs[t] : CHECKCB(t, cb) = true) do
85:         CO_DELIVER(m), m ∈ msgs: m.s = s, m.t = t, and m.c = c
86:         not_delvs[t] ← not_delvs[t] ∖ {⟨s, c, cb⟩}
87:         msgs ← msgs ∖ {m}
88:         last_delvs[t] ← last_delvs[t] ∖ {⟨s, _⟩} ∪ {⟨s, c⟩}
89:         causal_barrier[t] ← causal_barrier[t] ∖ cb ∪ {⟨s, c⟩}

90: function CHECKCB(topic t, causal barrier cb)
91:     for all ⟨s, c⟩ ∈ cb do
92:         if ( (∃ ⟨s', c'⟩ ∈ last_delvs[t]: s = s' and c' ≥ c)
              or (∃ ⟨s', c'⟩ ∈ first_rec[t]: s = s' and c' > c) ) then
93:             cb ← cb ∖ {⟨s, c⟩}
94:     return (cb = ∅)

95: procedure SENDACKS(j, m)
96:     if (⟨i, SUB, _⟩ ∈ view[m.t] and ∄⟨m.s, _⟩ ∈ first_rec[m.t]) then
97:         m.data ← m.data ∪ {⟨i, SUB, c⟩ : ⟨i, SUB, c⟩ ∈ view[m.t]}
98:     SEND(m) to $p_j$
```

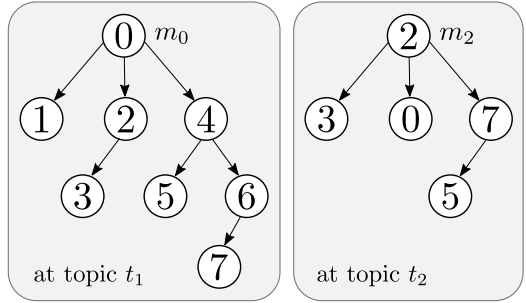

Figure 3: Broadcast trees for two different sources and topics.

the case of $p_4$ (CHILDREN$(4, t_1, 2) = \{5, 6\}$), it forwards $m_0$ to its children $p_5 \in c_{4,1}$ and $p_6 \in c_{4,2}$. Finally, $p_6$ sends $m_0$ to $p_7$.

Consider now a second example, on the right side of Figure 3, where only $p_0$, $p_2$, $p_3$, $p_5$, and $p_7$ are subscribers of $t_2$ and $p_2$ publishes $m_2$ related to $t_2$. In this case, $p_2$ sends $m_2$ to each of its child of its $\log_2 N = 3$ clusters that are also subscribers of $t_2$: CHILDREN$(2, t_2, 3) = \{3, 0, 7\}$ ($p_6$ is the first node in $c_{2,3}$ but it has not subscribed to $t_2$). Upon receiving $m_0$, $p_3$ does not forward it, because it is already a leaf node in the tree. Node $p_0$ does not forward it to $p_1$ since the latter is not a subscriber of $t_2$. On the other hand, $p_7$ verifies that in cluster $c_{7,2} = (5, 4)$, $p_5$ is a subscriber of $t_2$ (CHILDREN$(7, t_2, 2) = \{5\}$), and therefore sends $m_2$ to $p_5$ which on its turn does not send it to $p_4$, because even if $p_4$ is the first and only node in $c_{5,1}$, it is not a subscriber of $t_2$. For more details about how to build spanning trees over VCube, see [18].

After forwarding a message $m$ to a child $k$, node $i$ waits for an ACK message from $k$, which confirms the reception and propagation of $m$ by $k$. A node will send an ACK to its parent node only after it receives itself $ACK$ messages from all its current children related to the topic in question (lines 58-61). ACK messages will, thus, be propagated to the root, the source node of $m$. Eventually the latter receives all the $ACK$ messages it waits for and, in this case, the task related to $t$ removes the next message to be published from the queue associated to the topic $t$, if there is one. These sequences of SUB, UNS, or PUB and then ACK messages from/to the source ensure the per-source FIFO reception order of published messages of the topic.

**Reception and Delivery of Messages**: When receiving a PUB message $m$ of topic $t$ from $s$ (lines 63-68), if node $i$ is a subscriber of $t$ and has not already delivered $m$, it keeps $m$ in set $msgs$ and both its identification and causal barrier in set $not\_delvs[t]$. If $m$ is the first message received from $s$ to $t$, $i$ registers it in $first\_rec[t]$ in order to enforce the causal dependencies even under the dynamics of subscriptions. Then, node $i$ verifies, based on direct causal dependencies, which of the previously received messages can be delivered to the application. To this end, node $i$ invokes the function CHECKDELIVERY($t$) (lines 83-89) which, in it its turn, calls CHECKCB($t, cb$) in order to check direct dependencies (line 90-94). A message $m$ can be delivered to $i$ only when every message $m'$ on which $m$ causally depends either has already been delivered to $i$ or will never be received by $i$ because *VCube-PS* has not considered $i$ as a subscriber of $t$ during the construction of the spanning tree that broadcast $m'$. In other words, the first PUB message received from $s$ on topic $t$ by $i$ has a higher sequence number than the sequence number of $m'$. Such a detection of the first message is possible thanks to the $first\_rec_i[t]$ set and the fact that, for the same source, publications of messages of the same topic respect per-source FIFO order.

After deliver $m$, node $i$ removes it from its pending messages (lines 85-88) and updates its local causal barrier variable (line 89). Note that, since the delivery of one message $m$ can enable the delivery of other messages that causally depend on $m$, all remaining non delivered messages are rechecked.

**Membership Management**: In *VCube-PS*, distributed spanning trees are also used to notify membership changes. When a node $i$ subscribes (resp., unsubscribes) to (resp., from) a topic $t$, a broadcast SUB (resp., UNS) message will be received by all (resp., current subscribers) of $t$. Upon receiving either a SUB our UNS message, a subscriber of $t$ updates its view of the membership related to $t$ (line 72) by calling function UPDATE($set_1, set_2$) (lines 75-82) which merges two membership sets, keeping only the current subscribers.

When a node $i$ subscribes to a topic $t$, the ACK messages related to the SUB messages will also gather information about $t$'s membership. Function SENDACKS (lines 95-98) is responsible for sending ACK messages. Before forwarding a received ACK message to its parent, each subscriber of $t$ includes in the message its current view of $t$'s membership (line 97) merged with the partial membership information coming from its own children (line 57). When receiving all ACK messages from its children, the new subscriber $i$ is aware of $t$'s membership.

If node $i$ unsubscribes from topic $t$, it no longer delivers messages related to the topic (line 33). On the other hand, $i$ can continue to forward messages related to $t$ to the other subscribers of $t$ in the spanning tree if one of the following situations occurs: (1) there exist subscribers of $t$ that are not aware of $i$'s unsubscription, i.e., they have not received the corresponding UNS message from $i$ yet or (2) there are messages queued in $i$'s *br_queue[t]* waiting to be forwarded. Node $i$ also sends ACK messages to its parent node in the respective spanning tree. These ACK messages are related to published messages that $i$ received and forwarded before leaving $t$ or to messages that satisfy the above-mentioned situations. However, eventually all ACK messages will be sent and, thereafter, node $i$ will no more take part in the broadcast of messages related to $t$. When a subscriber of $t$ receives an UNS message related to node $i$, it removes $i$ from its view of $t$'s membership (line 72) as well as the information about the first message received from $i$ with regard to $t$ (line 74). The latter will be renewed if $i$ rejoins $t$ later.

*3.4. Proof*

A VCube of $N = 2^k$ is composed by two "sub-VCubes" of $n = 2^{k-1}$ connected by $2^k$ links. We define SubVC(i,k) as the set of nodes that belong to one of the hypercubes of a VCube of size $N = 2^k$ which contains $i$. For instance, in Figure 1, $SubVC(0,3) = SubVC(1,3) = \{0, 1, 2, 3\}$ while $SubVC(5,3) = SubVC(6,3) = \{4, 5, 6, 7\}$.

Let $\Delta_p^{m,t,k}$ be the time interval taken to propagate $m$ over the spanning tree induced by the broadcast of $m$ over a system of $N = 2^k$ nodes, logically organized in a VCube.

Let $m$ be a message related to topic $t$ broadcast by $s$. If $m$ is of type $PUB$ or $UNS$, we define $Sub_p^{m,t,k}$ as the set of nodes composed by $s$ plus the nodes which are subscribers of topic $t$ and the nodes of their respective "sub-VCube", which are subscribers of $t$ too, are aware of their subscription during $\Delta_p^{m,t,k}$, i.e., $Sub_p^{m,t,k} = \{i \mid \{\langle i, SUB, \_\rangle, \langle j, SUB, \_\rangle\} \subset view_j[t], \forall j \in \{SubVC(i,k) \cup \{s\}\}, \forall\, tm \in \Delta_m^p\}$. On the other hand, if $m$ is of type $SUB$, $Sub_p^{m,t,k} = \Pi$.

In the case of types $PUB$ and $UNS$, $Sub_p^{m,t,k}$ corresponds to those subscribers whose respective subscription to $t$ is known by a sufficient (not necessary) set of subscribers of $t$ which ensures the reception of $m$ by the former. Note that, except from node $s$, a node never sends a message to a node that does not belong to its own "sub-VCube" in the corresponding subtree.

**Lemma 1.** *Let $m$ be the first message of $queue\_br_s[t]$ of node $s$ related to topic $t$. Every node in $Sub_p^{m,t,k} - \{s\}$ receives $m$.*

*Proof.* Task $START\_MSG\_PROPAGATION(t)$ of $s$ removes $m$ from $queue\_br_s[t]$ and starts its propagation (line 15). The proof is by induction. For the induction basis $n = 2$, we consider $s = 0$ as the source of $m$ and node $1 \in c_{0,1}$. If $m.type = SUB$ or $\langle 1, SUB, \_\rangle \in view_0[t]$, task $START\_MSG\_PROPAGATION(t)$ of node 0 sends $m$ directly to node 1 which, therefore, receives $m$ ($Sub_p^{m,t,k} = \Pi$); otherwise $m$ is not sent to node 1 ($Sub_p^{m,t,k} = \{0\}$). Thus, the lemma holds. We consider now $N = 2^k$ and $N = 2^{k+1}$ as the induction hypothesis and step respectively. A VCube, of $2^{k+1}$ is composed by two "sub-VCubes" of $2^k$ connected by $2^k$ links. Let $s$ and $j$ be two of these nodes, respective roots of the the spanning trees built over the "sub-VCubes". In line 53, task $START\_MSG\_PROPAGATION(t)$ of $s$ sends $m$ to at most $k$ nodes, being $j = \text{FirstChild}(s,t,k)$ or the first node in $c_{j,k}$, included in $view_s[t]$. Thus, $m$ is propagated through the two distinct subtrees of size $N = 2^k$ and, based on the induction hypothesis, nodes of the subtrees receive $m$. □

**Lemma 2.** *$\Delta_m^p$ is finite.*

*Proof.* By Lemma 1, $m$ is only propagate through a spanning tree whose size is as most $\log_2 N$. □

**Lemma 3.** *Every message $m$ in $br\_queue[t]$ of $s$ is eventually broadcast to all nodes of $Sub_p^{m,t,k} - \{s\}$.*

*Proof.* Task $START\_MSG\_PROPAGATION(t)$ of $s$ always removes the first message from $br\_queue[t]$ (line 15) and, by Lemma 1 propagates it to the nodes of $Sub_p^{m,t,k} - \{s\}$. Every reception of $m$ by nodes of the $Sub_p^{m,t,k}$ is acknowledged (*ack* message), in a sink tree way, from the leaves to the root $s$ (lines 49 and 61). Hence, when $START\_MSG\_PROPAGATION(t)$ of $s$ receives all the acknowledge from its children it unblocks (line 31) and, if it exists, removes the first message from $br\_queue[t]$ and starts the propagation of it. Therefore, every message in $br\_queue[t]$ is eventually broadcast. □

**Lemma 4.** *If a node $s$ is a subscriber (resp., not a subscriber) of topic $t$, eventually all nodes are aware of it.*

*Proof.* When a node $s$ subscribes (resp., unsubscribes) to (resp., from) topic $t$, it calls function SUBSCRIBE($t$) (resp. UNSUBSCRIBE($t$)) that, in its turn, calls procedure CO_BROADCAST which includes a message $m$ of type $SUB$ (resp. $UNS$) in $br\_queue[t]$ 12. By Lemmas 1 and 3, $m$ will be received by every node $j$ in $Sub_p^{m,t,k} - \{s\}$. If $m$ is of type $SUB$, $Sub_p^{m,t,k} = \Pi$ (lines 24 and 42). Upon reception of $m$, $j$ includes (resp., remove) $s$ to (resp., from) $view_j[t]$ (line 72). If $type = SUB$, $s$ will be aware of the subscribers of $t$ since this information is included in the *ack* messages. Therefore, eventually every node will be aware that $s$ is a subscriber (resp., not a subscriber) of $t$, i.e., $\langle s, SUB, \_\rangle \in view_j[t]$ (resp., $\langle s, SUB, \_\rangle \notin view_j[t]$) $\forall j, 1 \leq j \leq N$). If $type = UNS$, due to Lemma 1, $s$ will no more receive any broadcast message related to $t$ till it subscribes to $t$.

□

**Lemma 5.** *(Integrity) Every node, subscriber of $t$, delivers at most once a published message $m$ and only if $m$ was previously published by some node.*

*Proof.* Let $s$ be the publisher of message $m$ related to topic $t$. Node $s$ calls function PUBLISH($t,m$) which, in its turn, calls procedure CO_BROADCAST that uniquely identifies $m$ and includes it with type $PUB$ in $br\_queue[t]$ (line 12). When task $START\_MSG\_PROPAGATION$ of $s$ handles the propagation of $m$, it delivers $m$ to $s$ by calling CO_DELIVER($m$) (line 19). As $m$ is removed from br_queue[t] of $s$, it will not be handled twice by the task. By Lemmas 1 and 3, $m$ is published and received by all nodes in $Sub_p^{m,t,k} - \{s\}$. The reception of this message by $j$ is only handled by task $HANDLE\_RECEIVED\_MSG$. If $s$ is not a subscriber of $t$, the task does not deliver $m$ since the condition of line 62 is false. Otherwise, $m$ and its identity are respectively included in $msgs$ and $not\_delvs[t]$, and, when $m$'s causal precedences are satisfied (lines 66 - 68 ), $m$ is delivered by procedure

CO_DELIVER($m$). It is delivered only once since CO_DELIVER($m$) only handles messages whose identities belong to $not\_delvs[t]$ and, after delivering $m$, $m$'s identity is removed from $not\_delvs[t]$, included in $last\_delvs[t]$, and $m$ removed from $msgs$ (lines 86-88). □

**Lemma 6.** *Let $m$ and $m'$, respectively identified by $\langle s, t, c \rangle$ and $\langle s', t, c' \rangle$, be two published messages such that $m \prec_{im} m'$ (direct dependence). Then $\langle s, c \rangle \in m'.cb$.*

*Proof.* By hypothesis $m \prec_{im} m'$, thus, either $s = s'$, i.e., $s'$ has published both messages or $s'$ has delivered $m$. In both case, $s'$ updates its $causal\_barrier_{s'}[t]$ variable (lines 22 and 89) which now includes $\langle s, c \rangle$. Furthermore, $s'$ does not delete $\langle s, c \rangle$ from $casual\_barrier_{s'}[t]$ until $m$ is sent (line 21). Consequently $\langle s, c \rangle$ will belong to $m'.cb$. □

**Theorem 1.** *(Safety Property): The algorithms ensure causal ordering of publishing messages for topic $t$.*

*Proof.* The formal proof of the safety property is inspired in the ones presented in [15] and [19]. Let $m$ and $m'$, respectively identified by $\langle s, t, c \rangle$ and $\langle s', t, c' \rangle$, be two published messages such that $m \prec_{im} m'$ (direct dependence) received by node $j$. Thus, we can infer by Lemma 6 that $\langle s, c \rangle \in causal\_barrier_j[t]$ and, then, procedures CHECKDELIVERY($t$) (lines 83-89) and CHECKCB($t, cb$) (line 90) ensure that $m'$ is delivered by $j$ provided that $m$ has been delivered or $m'$ is the first PUB message received by $j$ (line 94). Transitivity of message causality guarantees the causal order delivery of all messages. Lemma 5 ensures that messages are delivered only once. □

We define $\Delta_d^{m,j}$ as the time interval between the reception of $m$ by node $j$ and the deliver of $m$ by $j$

**Lemma 7.** *Let $m$ be a PUB message related to $t$, received by $j$. $\Delta_d^{m,j}$ is finite.*

*Proof.* The proof is by contradiction. The broadcasts of the $PUB$ messages, which precede $m$, that were received but not delivered by $j$ define a partial order. Let $m'$, identified by $\langle s, c \rangle$, be a message in this partial order whose broadcast does not have any predecessor. Since, $m'$ has not been delivered by $j$, $\langle s, c, \_ \rangle\} \in not\_delvs$ and $\langle s, c, \_ \rangle\} \notin last\_delvs$. This implies that there exists a message $m''$ sent to $j$ such that the publishing of $m''$ precedes the publishing of $m'$ which violates the assumption that $m'$ has no predecessor. □

**Theorem 2.** *(Liveness Property): The algorithms ensure that every published message m related to t is eventually delivered by all j in $Sub_p^{m,t,k}$, provided that j has not unsubscribed from t during $\Delta_d^{m,j}$.*

*Proof.* The proof is inspired by the one in [15]. Due to Lemmas 1 and 3, $j \in Sub_p^{m,t,k} - \{s\}$ receives receive $m$. By Lemma 6, $m.cb$ carries $\langle s', c' \rangle$, the identity of message $m'$, only if $m \prec_{im} m'$. Thus, $m.cb$ does not include any spurious information that could prevent the delivery of $m$ by $j$. When task $HANDLE\_RECEIVED\_MSG$ of $j$ receives $m$, it can deliver $m$ when CHECKCB$(t, cb)$ returns true, i.e., all causal precedences of $m$ have been satisfied. Thus, the delivery of $m$ can only be delayed due to other PUB messages related to $t$, that precede $m$ which have not been delivered yet. By Lemma 7, this delay, $\Delta_d^{m,j}$, is finite. Therefore, $m$ is eventually delivered. □

## 4. Experimental Results

In order to assess the performance of *VCube-PS* with different configuration scenarios, we conducted experiments on top of the event-driven PeerSim [14] simulator. In the majority of scenarios, we compare *VCube-PS* to *SRPT*. For each topic, *SRPT* selects a node to act as the root of the broadcast tree for the respective topic.

In the experiments, we consider that each message exchanged between two nodes consumes $t_{pc} + t_q + t_t + t_{pp} + t_d$ units of time (*u.t.*). Apart from $t_d$ which represents the time necessary for a subscriber to satisfy all causal dependencies before delivering a message to the application, all other components are based on a packet-switched network delay model [20]: $t_{pc}$ accounts for the processing time of a message by a node, e.g., checksum verification and routing decisions; $t_q$ is the time a message must wait in the queue before being transmitted; $t_t$ is the time necessary to transmit all bits of the message into the link, and $t_{pp}$ expresses how long it takes for a message to transverse the link and reach the next hop. Assuming that there is no broadcast feature available in the system, if a message is sent to multiple destinations, a copy of the message is queued for each of the destinations. For our experiments, the ratio between $t_{pc}$ and $t_{pp}$ has an impact on the threshold value for starting to queue messages as well as how fast the queue grows. Hence, based on [21], we set $t_{pc} = t_t = 1$ *u.t.* and $t_{pp} = 100$ *u.t.* (1/100 ratio).

For most experiments, the number of nodes $N$ varies from 8 up to 4096, in a power of two, and each experiment was executed 40 times.

We consider the following metrics for comparison: (1) *Latency*: the time that a published message takes to be received and delivered by all subscribers; (2) *Number of messages*: overall number of PUB messages; (3) *Number of messages to be processed by a node*: size of the queue of each node; (4) *Size of PUB messages*: characterizes the number of direct causal dependencies that PUB messages hold; and (5) *Number of false positives*: number of messages received by nodes that act as forwarders of messages of type PUB.

*4.1. A Single Publisher*

This experiment evaluates the impact of the logarithmic properties of *VCube-PS*. A single publisher publishes one message. Hence, when a subscriber receives the message, there is no delay for delivery. Figure 4(a) shows the delivery latency when the number of nodes of the system varies and either 25% or 100% of them are subscribers. The set of subscribers is randomly chosen following a uniform distribution. In the case of 4096 nodes with 25% of subscribers uniformly distributed, latency in *VCube-PS* is on average 533 units of time, 26% less compared to the one presented by *SRPT* in the same scenario (720 *u.t.*) We remark that when 100% of the nodes are subscribers, *SRPT* has no forwarder and, therefore, the latency of both Pub/Sub systems is always proportional to $\log_2 N$. The only difference in this case is that *SRPT* has an additional hop as the message to be published must be sent to the root of the single tree.

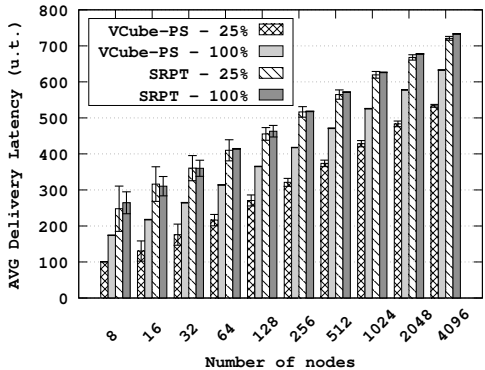 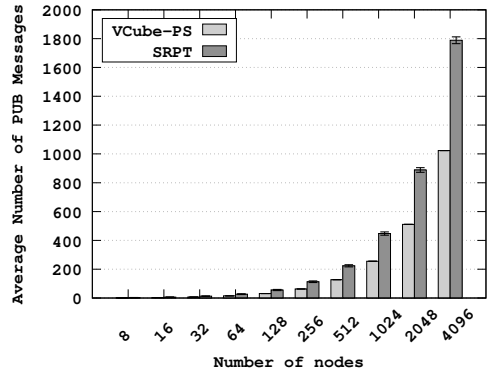

(a) Average latency.  (b) PUB messages with 25% of subscribers.

Figure 4: Average latency and number of PUB messages for *VCube-PS* and *SRPT* with different number of subscribers.

The average number of PUB messages follows the same behavior as shown in Figure 4(b). In the figure, for the two approaches with 25% of the nodes as subscribers, *VCube-PS* always presents the same number for PUB messages, since there is no forwarder in the tree. On the other hand, forwarders in *SRPT* are responsible for up to 2.7 times more messages (for 8 nodes) compared to *VCube-PS*. As the number of nodes increases, this difference is reduced, although *VCube-PS* generates, on average, at least 43% fewer messages than *SRPT* (4096 nodes).

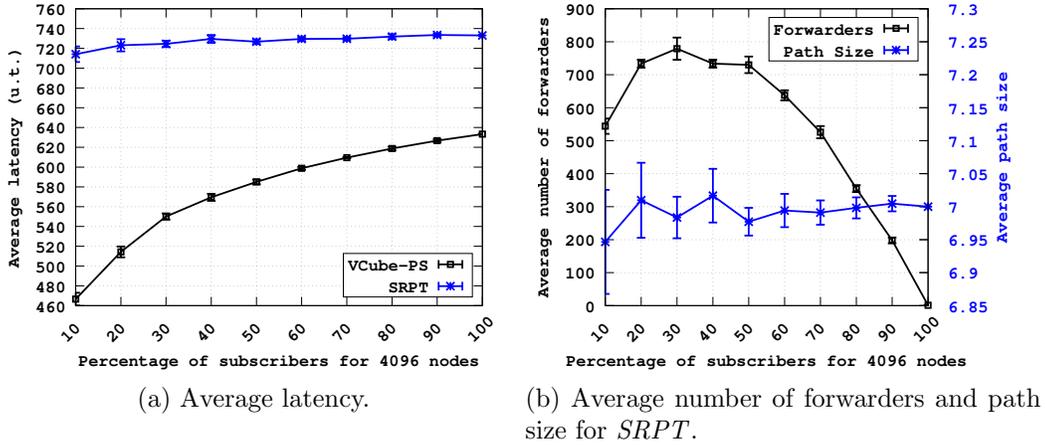

(a) Average latency.

(b) Average number of forwarders and path size for *SRPT*.

Figure 5: The impact of different numbers of subscribers on a network with 8192 nodes.

Figure 5 provides a more detailed analysis of the impact of the number of subscribers in *VCube-PS* and *SRPT* performance. The system has 4096 nodes and the number of subscribers, uniformly distributed, varies from 10% up to 100%. In Figure 5(a), we observe that *VCube-PS* performs logarithmically with respect to the number of subscribers while *SRPT* does not. A tenfold increase of the number of subscribers induces just a 36% increase of the average latency of *VCube-PS*. On the other hand, even if the average latency of *SRPT* varies up to approximately only 2.7%, it is always higher compared to *VCube-PS*.

Figure 5(b) helps to better understand the behavior of *SRPT*. If a subscriber is a leaf node, the tree will have a branch with 12 levels, even if no other node in the branch is a subscriber. When 30% of the nodes are subscribers (i.e., around 1228 nodes), there exist, on average, 779 forwarders, resulting in a tree with almost 50% of the nodes of the system. However, as the number of subscribers increases, they replace forwarders in the tree. Despite of this behavior, due to the uniform distribution of subscribers, the average path size that the message travels over the tree follows a constant pattern (around 7 hops) no matter the percentage of subscribers.

4.2. Several publishers

In these experiments, all nodes are subscribers of a single topic and the number of publishers varies. Each publisher $i$ sends one message at time $t_i$ which is uniformly distributed between $[0, 1000]$ units of time. By having multiple publishers of the same topic, differences in latency will arise from the distribution of the load among the nodes when using one root per publisher (*VCube-PS*) or one root per topic (*SRPT*).

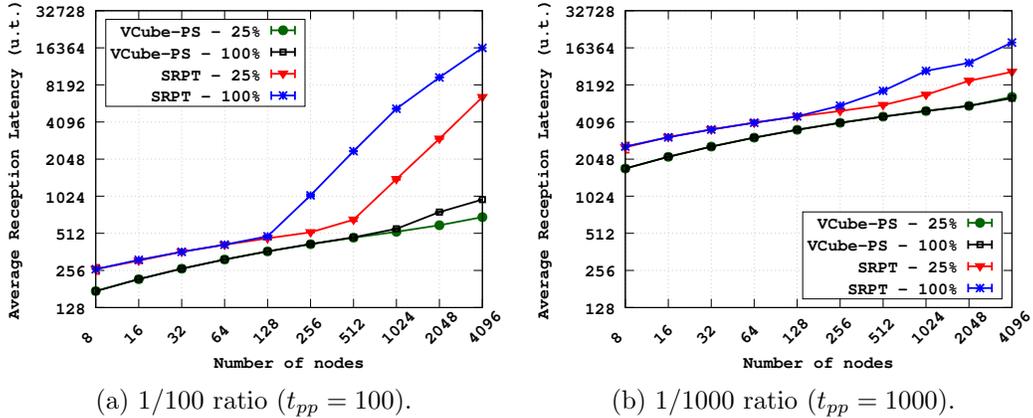

(a) 1/100 ratio ($t_{pp} = 100$).     (b) 1/1000 ratio ($t_{pp} = 1000$).

Figure 6: Average reception latency with 25% and 100% of publishers (log. scale).

Figure 6 shows in logarithmic scale the average reception latency when the number of nodes of the system varies and either 25% or 100% of them are publishers. Since the ratio between the processing time ($t_{pc}$) and the propagation time ($t_{pp}$) has an impact on the load contention, we consider the ratio 1/100 (Figure 6(a)), which is used in all other evaluations of this work, but also a propagation time which is ten times greater, ($t_{pp} = 1000$ *u.t.*), leading to a ratio 1/1000 (Figure 6(b)).

We can observe in Figure 6(a) that *VCube-PS* has a load distribution with a maximum increase of 38.8% (4096 nodes and 100% of publishers)

Table 1: Average size of the queue per group of nodes.

| # of messages | # of nodes (VCube-PS) | # of nodes (SRPT) |
|---|---|---|
| 0 | 0 | 512 |
| (0, 2] | 0 | 448 |
| (2, 4] | 0 | 60 |
| (4, 8] | 495 | 3 |
| (8, 16] | 510 | 0 |
| (16, 32] | 19 | 0 |
| (32, 4096] | 0 | 0 |
| (4096, 8192] | 0 | 1 |

when compared to *VCube-PS* with 4096 nodes and 25% of publishers. It happens because even though there are 4 times more messages, they traverse different paths in the network. On the other hand, in *SRPT*, if several messages arrive at the root of the tree at the same time they will be queued before transmission, increasing, thus, the reception latency. For up to 128 nodes, *SRPT* latencies are on average one hop in time higher compared to *VCube-PS*, because in these cases the arrival and output rates of messages are close, leading to no contention. Beyond this number of nodes, the root receives more messages than it can process and transmit per interval of time and starts to saturate. For instance, in comparison with *VCube-PS* with 256 nodes and 100% of publishers, *SRPT* has an average latency 2.48 times greater, and this ratio grows linearly after this point.

Comparing Figure 6(b) to Figure 6(a), the average reception latency increases less in *SRPT* in relation to *VCube-PS* because, with a 1/1000 ratio, it takes longer to receive messages, although the output throughput remains the same.

Table 1 shows the distribution of nodes according to the average size of their sending queues, in a scenario with 1024 nodes, 1/100 ratio, and where all nodes are publishers and subscribers.

The load distribution on the nodes in *SRPT* is uneven when compared to *VCube-PS*: 98% of the nodes in *VCube-PS* have an average load between (4, 16] messages, while 44% of the nodes in *SRPT* have on average between (0, 2] messages in their buffers. In *SRPT*, 50% of the nodes simply do not participate in the routing of any message, because they are leaf nodes of the single tree of the topic and one node (the root) has an average load of 9240 ($\sigma = 4617$) messages, which incurs in high reception latencies.

*4.3. Message Order*

Besides the published message itself, every PUB message contains its causal barrier, i.e., a list with direct causal dependencies of the published message. Thus, the size of a PUB message increases depending on the number of elements in this list. In order to evaluate the size of such a list and the latency due to message ordering, we consider that one node $s$, chosen randomly, publishes a first message $m_s$. Upon receiving it, each node $k$ waits for a random time ($t_w$) before broadcasting message $m_k$, similarly to a message discussion group service where all members of the group answer publicly to a question posted by one of them. For $N$ nodes, there will be $N^2 - N$ messages. Additionally, we extend this scenario for the case when a node $k$ has to wait for at least $p$ messages before broadcasting its own. To this end, there are $p \geq 1$ nodes that independently broadcast a message, each in the beginning of the experiment. Just after receiving all these initial messages, any node can publish a message.

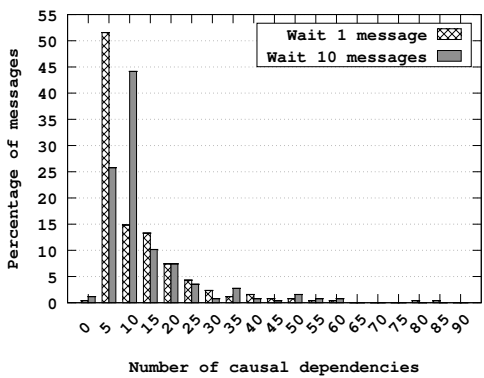

Figure 7: Frequency distribution for the number of causal dependencies of a message in a network running *VCube-PS* with 256 nodes.

Figure 7 groups messages according to the size interval of their causal barriers for *VCube-PS*. When it is necessary to wait for just one message before a node broadcasts its own message, 51.6% of the messages generated in the system have less than 5 preceding messages. More precisely, 19.9% of them have just one causal dependency. On the other hand, if a node waits for more messages (10 in the case of the figure) before broadcasting its own, a larger number of nodes will have 10 or more direct dependencies. In this case, 35.2% of the messages have size 10 (10 direct dependencies) and 79.7% of them have fewer than 15. However, in both cases, the number of direct dependencies keeps a reasonable size.

We also evaluated the additional delay imposed by causal barriers before delivering a message to the application. When a node waits for 1 message before broadcasting its own, about 95.1% of the messages are delivered in less than 10 $u.t.$ after the message is received (87.2% are delivered with no delay). Only 81 messages (out of 65280) have a delay higher than 50 $u.t.$, with an upper limit of 150 units of time. Increasing the number of the waiting messages to 10, 457 messages wait more than 50 $u.t.$ to be delivered (maximum 187), although the number of messages with no delay remains high (84.2%).

*4.4. Multiple Topics*

As discussed in [9], in real world applications like Twitter, a few topics are related to most of the messages. The authors show that in Twitter, roughly 60% of the topics have only one message published, 83% of them have no more than 5, only 0.15% of the topics are related to more than 1000 messages each. This behavior follows a Zipf-like distribution with a coefficient of 0.825 according to the data provided in the reference. We evaluated *VCube-PS* and *SRPT* with multiple topics. Messages are assigned following both the Zipf-like and uniform distributions. Figure 8 depicts the results for 256 nodes, 128 topics, and a varying number of messages. Each node publishes a new message on average every 500 $u.t.$ for a topic, randomly chosen. Therefore, messages are uniformly distributed among the publishers, but not necessarily among the topics.

No matter the distribution of messages among the topics, *VCube-PS* always relies on the same root for a given publisher, while *SRPT* does not. This is the reason why the behavior of *SRPT* is the same as *VCube-PS*'s for a uniform distribution of messages. However, when the number of messages sent per node increases beyond a threshold, *VCube-PS* increases the latency due to contention at the source of the messages, i.e., the root of the tree. On the other hand, for the Zipf distribution, *SRPT* has an average reception latency 30.6% higher compared to the uniform distribution (for $2^{14}$ messages). *VCube-PS* increases latency, on average, only 9.2%.

These results confirm that *VCube-PS* is scalable in terms of publishers, while *SRPT* is scalable in terms of topics. However, in real scenarios, most of the messages are concentrated on a small number of topics.

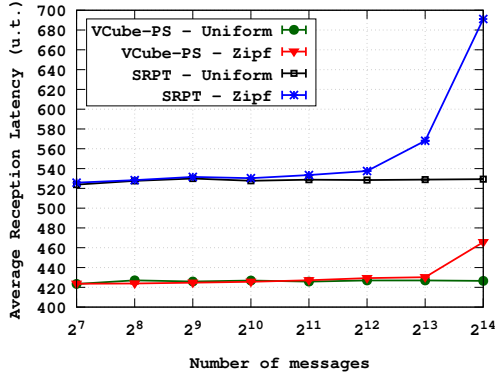

Figure 8: Average reception latency with 256 nodes and 128 topics for two distribution of messages per topic.

*4.5. Churn Evaluation*

In these experiments, we are interested in evaluating how *SRPT* and *VCube-PS* tolerate membership changes. In *VCube-PS*, these changes are broadcast to all nodes using multiple trees, while *SRPT* has to restructure its single tree. For a given number of nodes and a topic, initially 75% of the nodes are randomly chosen as subscribers of the topic. Every 300 units of time, 12.5% (resp., 25%) of the nodes unsubscribe from the topic and new ones join it. In the simulations, one random publisher sends a new message once every 500 units of time, up to a limit of 128 messages. Remember that *SRPT* often presents in its tree forwarders (non-subscriber nodes) and in *VCube-PS* when a node $i$ unsubscribes, it can still receive and send publications related to the topic for a while (temporary forwarders, see Section 3.3).

Figure 9(a) presents the average reception latency. We can observe the impact of the churn in *SRPT* average reception latency: the higher the number of nodes, the higher the latency, except for 256 where both approaches perform almost the same. Such an overhead is due to the variable number of forwarders, increasing number of message induced by tree restructuring, as well as higher root contention. In particularly, with 2048 and 12.5% churn rate (worst case), latency is about 30 times higher than *VCube-PS*'s. On the other hand, the same figure shows that churn does not induce such a degradation in reception latency in *VCube-PS* since trees are dynamically built at each broadcast. However, we should point out that, increasing churn rate from 12.5% to 25% leads to more temporary forwarders which increases

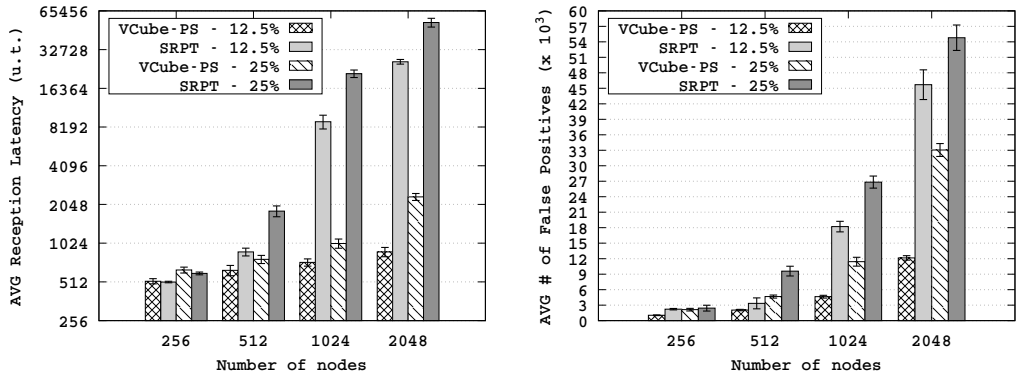

(a) Average reception latency (logarithmic scale).

(b) Average number of false positives.

Figure 9: Publication of 128 messages in different network configuration with 75% of subscribers and churn rate of 12.5% or 25%.

*VCube-PS* latency. For instance, with 2048 nodes, average reception latency in *VCube-PS* is 40.3% higher with 25% churn rate than with 12.5%.

Figure 9(b) shows the average number of messages received by forwarders (false-positives). Because of unnecessary forwarders, for both churn rates, *SRPT* always presents more forwarders compared to *VCube-PS*, reaching up to 3.7 times more false-positives than *VCube-PS* with 2048 nodes and 12.5% of churn rate (worst case). False-positives lead to 27.8% extra sent PUB messages in *SRPT* (2048 nodes, 25% churn), while for the same scenario, *VCube-PS* has only 16.8% extra messages due to temporary forwarders.

*4.6. Broker-based SRPT*

For the results presented in this section, we alsso consider a *SRPT* Pub/Sub system based on brokers (e.g., DYNATOPS [5], see Section 5). We denote it *SRPT*-B and the previous one, we renamed to *SRPT*-S. In *SRPT*-B, the single broadcast tree per topic is composed by nodes that are either brokers (instead of subscribers) or forwarders. Subscribers are directly connected to brokers, according to their locality and/or interests. Each published message for this topic is transmitted over this tree and each broker, upon reception, directly sends the message to the subscribers connected to it.

Figure 10 shows the average reception latency for *SRPT*-S, *SRPT*-B, and *VCube-PS*. Publishers are randmomly chosen among the subscribers of the topic and send a new message on average every 500 $u.t.$, up to a limit of 128 messages.

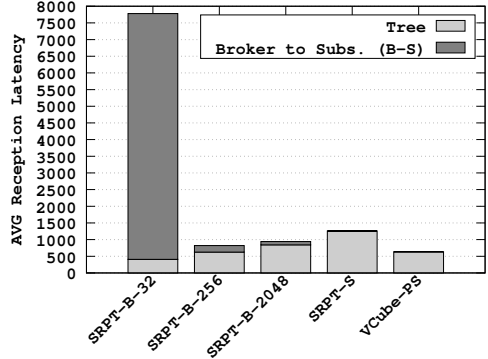

Figure 10: Average reception latency for different approaches and 4096 nodes.

We defined three configurations for $SRPT$-B with different number of brokers: 32, 256, and 2048. The other nodes are subscribers evenly distributed among the brokers: 127, 15, and 1 subscribers per broker respectively. Note that for both $SRPT$-B and $SRPT$-S, we considered no forwarders, i.e., trees are composed only by the respective numbers of brokers or by 4096 subscribers, respectively.

In $SRPT$-B, average reception latency is composed by the time to send the message to the brokers (`Tree` in the figure) plus the time for the broker to send the message to the connected subscribers (`B-S` in the figure). On the one hand, we observe in the figure that the fewer the number of brokers, the lower the `Tree` reception latency. On the other hand, the fewer the number of brokers, the higher the number of messages per broker forwarded to the subscribers, inducing broker-level contention (like the root in $SRPT$) and, therefore, the higher the `B-S` reception latency. In the configuration with 32 brokers, average `B-S` latency is 4 times higher than the other $SRPT$-B's `B-S` latencies due to high broker-level contention while, with 256 nodes, the load is better distributed.

We also point out that even if *VCube-PS* builds trees with bigger height compared to $SRPT$-B's, it presents lower average reception latency than all $SRPT$-B configurations (22% better for $SRPT$-B with 256 brokers) since it avoids contention by exploiting multiple paths. A last observation is that $SRPT$-B with 2048 brokers has lower reception latency than $SRPT$-S since the latter presents more contention in the root of the tree, which is composed by 4096 subscribers.

## 5. Related Work

Basically, there exist two models of Pub/Sub systems: *topic-based* [3, 4, 6, 5] and *content-based* [12, 22]. In the first model, subscribers share a common knowledge on a set of available topics and every published message is labeled with one of these topics. In a topic-based Pub/Sub system, a subscriber can register its interest in one or more topics, and then it receives all published messages related to these topics (e.g., Scribe [3], Bayeux [4], DYNATOPS [5], and Dynamoth [6]). In the content-based model [23], messages are structured based on multiple attributes, and subscribers express their interests by specifying constraints over the values of these attributes (e.g., SIENA [24], JEDI [12] and BlueDove [25]). The content-based model provides more flexibility to subscribers for defining their interests, but at the expense of more complex user interfaces and the need for filtering. On the other hand, topic-based systems provide simpler and more efficient implementations and they are usually deployed in contexts where efficient and fast notifications are required.

Similarly to *VCube-PS*, many Pub/Sub systems use tree-based overlays (e.g., Scribe [3], Bayeux [4], Marshmallow [26], DR-Tree [22], and DYNATOPS [5], Magnet [7], DRScribe [8]). The advantage of using trees is the logarithmic guarantees on publication reception time with respect to the number of nodes that compose a tree. However, contrarily from *VCube-PS*, most solutions often implement one single multicast tree (usually one per topic in topic-based systems), statically constructed from the start or as nodes join the system. Consequently, every publication should be broadcast from the root of this tree that might, then, become a bottleneck. Moreover, many of these multicast trees include unrelated intermediate hops and nodes that are not subscribers which have to forward the message presenting, thus, the problem of false positives and the need of message filtering (e.g. DR-Tree [22] and Scribe [3]). Finally, the maintenance cost is usually high, specially in presence of churn.

Several solutions (e.g. Scribe [3], DYNATOPS [5], Magnet [7], DRScribe [8], etc.) construct independent multicast trees on top of Distributed Hash Table (DHT) overlays (e.g Pastry, CAN). They adopt the *rendezvous* point approach, where a node, responsible for the hashed key of a topic name, becomes the *rendezvous* point, i.e., the root of the multicast related to the topic. In order to join this tree, a node seeks a DHT path that leads to the root. Thus, nodes in the tree are either subscribers/brokers of the topic or merely forwarders, which are added to the tree because they are in the path towards the root. Some DHT overlay like PeerCube [27] and HOMED [28] are based on a hypercube-like topology themselves.

Few Pub/Sub systems ensure message ordering [29, 12, 30, 31], in this case, usually total order. Authors in [29] propose a top-basic Pub/Sub system where messages published on different topics are either delivered in the same order to all subscribers or tagged as out-of-order (*weak total order*); while in [30], the task of ordering messages is distributed across sequencer nodes which totally order messages for the same topic. Considering FIFO links, [31] presents a distributed total order protocol for a content-based Pub/Sub system where a broker can decide if a message can be delivered immediately or a consistent delivery order is required. In [32], the variations of end-to-end delay of messages, directly related to out-of-order FIFO delivery, are measured. Based on such analysis, nodes delay or not the delivery of a message aiming at reducing FIFO delivery order violations. JEDI [12] is a Pub/Sub system that ensures causal order. The latter is implemented by using a *return value*, a message for the receiver to notify the producer that a message was delivered, unlike *VCube-PS*, which does not require these extra messages since causal dependencies of a message are included in the message itself (*causal barriers*).

For the sake of scalability, several works have proposed the implementation of distributed causal order broadcast algorithms using causal barriers [33, 19, 15].

*VCube-PS* relies on a hypercube topology to built its spanning trees. Works like the one presented by [27] also take advantage of the low degree and path diameter of the hypercube topology to cope with the impact of high node churn in dynamic environments. However, the latter is not a Publish/subscriber system. HOMED is a content-based Pub/System proposed in [28] that maps nodes to a logical hypercube.

## 6. Conclusion

In this work we presented *VCube-PS*, a distributed topic-based Pub/Sub system. *VCube-PS* propagates information about membership changes and disseminate publish message to the subscribers of a topic using dynamic spanning trees built on top of a hypercube-like topology that presents multiple logarithmic features and is scalable by definition. While most other Pub/Sub approaches use static trees and *rendezvous* points, *VCube-PS* creates a new spanning tree rooted on the source of every message that is published, without any extra cost due to *VCube*'s properties. As the spanning trees contain only subscribers of some particular topic, the trees have a shorter height when compared to a per-topic single root tree and, therefore, lower latencies and number of messages. Furthermore, *VCube-PS* enforces the causal delivery of messages using causal barriers adapted to cope with the dynamics of the system.

Experimental results from simulations on PeerSim confirm the logarithmic properties of *VCube-PS*. Compared to an approach with one single root per topic, our solution presents the best results under a high publication rate per topic since it intrinsically provides load balancing. Furthermore, *VCube-PS* does not employ permanent forwarders which induce false positives and employs decentralized message broadcast which is efficient in terms of time.

Future directions of our work include adapting the proposed strategy to tolerate node faults. Furthermore, we envision the causal aggregation of messages that follow the same routes in order to reduce the number of messages employed without degrading the latency.